\documentclass[aps]{revtex4-2}
\usepackage{longtable}
\usepackage{graphicx}% Include figure files
\usepackage{dcolumn}% Align table columns on decimal point
\usepackage{bm}% bold math

\begin{document}
\title{P-V relationship of elements in the pressure range of 200--300 GPa}
%\textbf{P-V relationship of elements in the multi-megabar pressure
%Yuichi Akahama\textsuperscript{1} and Masaaki Geshi\textsuperscript{2}
%\textsuperscript{1}Graduate school of Science, University of Hyogo,
%3-2-1 Kamigohri, Hyogo 678-1297, Japan\\
%\textsuperscript{2}R3 Institute for Newly-Emerging Science Design, Osaka
%University,
%1-3, Machikaneyama, Toyonaka, Osaka, Japan
\author{Yuichi Akahama}
\author{Masaaki Geshi}
\affiliation{Graduate school of Science, University of Hyogo,
  3-2-1 Kamigohri, Hyogo 678-1297, Japan}
\affiliation{R$^3$ Institute for Newly-Emerging Science Design, Osaka University,
  1-2, Machikaneyama, Toyonaka, 560-0043, Japan}
\date{\today}
\begin{abstract}
In this study, we analyzed the pressure-volume (\emph{P}--\emph{V})
relationship of elements using the equation of state at an ambient
temperature within the multi-megabar pressure range of 200--300 GPa. We
investigated the compressibility of elements under ultra-high pressures
based on their positions in the periodic table. For the elemental
materials in this region, pressure (\emph{P}) can be approximated as an
\emph{N}-order function of volume
(\emph{V}): \emph{P} $\propto$ $1/V^N$.
By determining the \emph{N} value,
we can evaluate the contribution of the volume change to the total
energy of the system.
The \emph{N} value is also an indicator the bulk modulus in this pressure range
and shows periodicity with increasing atomic number, and shows significant
values in the range of 4.5--6, for transition metals with close-packed
structures (fcc, hcp). This suggests that the total energy of these
elements increases rapidly as volume decreases. Furthermore, the current
results provide basic data for understanding the compressibility of
elemental materials under extreme ultra high-pressure conditions based
on the quantum theory of solids.

\end{abstract}
\maketitle

\section{I. Introduction}

In recent years, the development of ultrahigh-pressure synchrotron X-ray
diffraction technique,\cite{ref1,ref2} combined with advancements
in first-principles calculation methods, has led to the elucidation of
structural phase transitions and equations of state (EOS) for nearly all
elemental materials under ultrahigh pressures. Using these experimental
techniques, we have clarified the structural phase transitions for
approximately 30 types of elements under ultrahigh-pressure conditions
as well as the EOS of each high-pressure phase.\cite{ref3,ref4,ref5,ref6,ref7,ref8,ref9}
These studies have revealed that nearly all elements exhibit metallic
behavior in the multi-megabar pressure range, often accompanied by
structural phase transitions.

Currently, EOSs for many elements at room temperature are proposed based
on pressure-volume (\emph{P}--\emph{V}) data obtained from X-ray
diffraction experiments in high-pressure regions. Pressure (\emph{P}) is
normally expressed as a function of volume: \emph{P} = F(\emph{V}). This
is commonly represented by the Birch--Murnaghan (BM)\cite{ref10}
equation and/or Vinet\cite{ref11} equation of (1) and (2), as
shown below:
\begin{equation}
P = (3/2) K_{0} ( x^{7/3} - x^{5/3} ) \cdot [1 +(3/4)(K'_{0} - 4)(x^{2/3} - 1)],
\end{equation}
and
\begin{equation}
P = 3 K_{0} [(1 - x^{1/3})/x^{2/3}] \cdot \exp{[3/2(K'_{0} - 1)(1 - x^{1/3})]},
\end{equation}
where $x = V/V_{0}$ and $K_{0}$ and $K'_{0}$ correspond to the bulk
modulus at normal pressure and the pressure derivative of
$K_{0}$, respectively. However, because both equations
express pressure as a complicated function of volume, understanding the
compression properties of these elements in the extreme high-pressure
region is challenging.

In this study, we provide an overview of the ultrahigh-pressure world of
materials by examining the atomic volumes and \emph{P}--\emph{V}
relationships of elements in the range of 200--300 GPa based on the
periodic table. Our findings show that in this ultrahigh-pressure range,
pressure can be approximated as a simple function of volume:
$P \propto 1/V^{N}$ with a parameter $N$. For
transition metals, \emph{N} has significant values in the range of
4.5--6, indicating that pressure has a strong dependence on volume. The
obtained results provide important information for the advancement of
not only high-pressure science but also materials science as well as
earth and planetary science.

\section{II. Analysis}

The atomic volumes ($V_{200}$ and $V_{300}$) at 200 and 300 GPa for
each element at an ambient temperature and the ratio of
$V_{300}/V_{\rm 0amb}$, where $V_{\rm 0amb}$ corresponds to the atomic volume of
the stable phase under ambient conditions, were estimated based on
the EOS obtained from the X-ray
diffraction experiments. The log(\emph{P})--log(\emph{V}) plot for each
element showed good linearity in the pressure range of 200 to 300 GPa.
Therefore, we performed a linear approximation ($\log(P) \propto -N\log(V)$)
in this region to estimate the coefficient \emph{N}. Fig. 1 shows the
$\log(P) - \log(V)$ plots for Ni,\cite{ref9} Pt,\cite{ref12,ref13,ref14} and
Bi.\cite{ref15} From these figures it is found that
$\log(P)$ and $\log(V)$ exhibit relatively good linearity in the
multi-megabar pressure range of 200--300 GPa. The standard deviation of
the \emph{N} values was within 0.5\%. In this pressure range, a simple
relational expression between \emph{P} and \emph{V} holds true,

\begin{equation}
  P \propto 1/V^{N}. 
\end{equation}

In this study, we also incorporated data from other studies. In cases
where the maximum pressure ($P_{\rm max}$) in measurements
did not reach 300 GPa, the reported EOSs were extrapolated. Some of
these cases had $P_{\rm max}$ value of less than 100 GPa.
The pressure scale used was the Pt--EOS developed by
Holmes \emph{et al}.,\cite{ref12} and the obtained \emph{N} value was
approximated 1\% smaller than that of the Pt--EOS recently revised by
Fratanduono \emph{et al}.\cite{ref14} By contrast, for the
Pt--EOS proposed by Dewaele \emph{et al}.,\cite{ref13} the
\emph{N} value was 5.5\% smaller. Furthermore, the
Ag\cite{ref16} data were recalculated using the Au scale
developed by Fratanduono \emph{et al}.\cite{ref14} In addition,
although the EOS proposed by other researchers were often used pressure
scales different from the Pt--EOS proposed by Holmes \emph{et al}.,
\cite{ref12} these EOSs were used in the analysis without pressure scale
correction.
The \emph{P}--\emph{V}
curves Pt--EOSs reported by Holmes \emph{et al}.,
Dewaele \emph{et al}., and Fratanduono \emph{et al}. and pressure
differences among them are shown in Fig. S1.

The EOS data for elements and the obtained results in this analysis are
listed in Table I, including atomic No., element symbol,
$P_{\rm max}$, atomic volume at ambient pressure
($V_{0}$), bulk modulus ($K_{0}$), pressure derivative of $K_{0} (K'_{0})$,
EOS formula: Vinet (V), Birch-Murnaghan (BM), and AP1 (polynomial, see Ref. 37),
atomic volumes at 200 GPa and 300 GPa ($V_{200}$ and $V_{300}$),
atomic volume of the ambient pressure phase under ambient
condition($V_{\rm 0amb}$), volume ratio ($V_{300}/ V_{\rm 0amb}$), crystal
structure/phase, and reference number.
The column of the structure/phase shows the high-pressure structure
or phase of the EOS based on the references. Some of EOSs are estimated
from P-V data of two phases (see the references for the structure
notation and its details).
Note that the EOSs for $\beta$--Po
type--S and hcp--Sn were calculated assuming the atomic volumes at 132
and 157 GPa to be $V_{0}$, respectively. The EOS of
bcc--Hf was recalculated using Vinet's equation,
assuming the atomic volume 15.32 Å\cite{ref3} at 60.9 GPa to be
$V_{0}$. The EOS of hcp--Be\cite{ref17} was
calculated using Vinet's formula.

\section{III. Results and Discussion}

Fig. 2 shows the changes in the atomic volume at 200 and 300 GPa
($V_{200}$ and $V_{300}$), the volume ratio $V_{300} /V_{\rm 0amb}$,
and the coefficient \emph{N} with respect to the atomic number. The
solid and broken vertical lines correspond to the periods of the
elements and boundaries between the transition elements and typical
elements or alkaline earth metals, respectively. Elements with
$P_{\rm max}$ higher than around 200 GPa are indicated in blue.
Note that by assuming \emph{N} = 4 in this pressure region,
the uncertainty in $V_{200}$ and $V_{300}$ is estimated to be 1.8\%,
even with a 10\% pressure uncertainty due to differences
in pressure scales; this falls within the symbol.
Estimating the error in the atomic volumes $V_{200}$ and $V_{300}$
obtained by (1) extrapolating the EOS estimated from $P$--$V$ data with
\emph{P}\textsubscript{max} of 100 GPa or less to the multi-megabar
region and (2) estimating from the EOS of elements that exhibit
structural phase transitions in the multi-megabar pressure region is
notably difficult. Therefore, these data should be considered reference
values. In particular, regarding the BM equation with
$P_{\rm max}$ of 100 GPa or less and $K'_{0}$ of approximately three less,
such as Be, Y, and Hf, we know that the $P$--$V$ curve
sometimes exhibits abnormal behavior when the equation is extrapolated
up to 250 GPa. Therefore, extrapolated $P$--$V$ curves for Be,
Y, and Hf were examined. Consequently, unnatural behavior was observed
for Hf (Fig. S2). The EOS of bcc--Hf was recalculated using the Vinet
formula. In addition, to evaluate the difference in the pressure scale,
the EOS of Ta obtained using the Pt--EOS scale proposed by Dewaele
\emph{et al}.,\cite{ref13} was recalculated using the Pt--EOS
scale proposed by Holmes \emph{et al}.\cite{ref12} The
recalculated results for Hf, Ta, and Pt have been added to Table I and
are shown in Fig. 2 with red symbols.

Under such ultrahigh pressures, the atomic volumes
$V_{200}$ and $V_{300}$ exhibit considerably smaller values than
those at normal pressure,\cite{ref18} and the range of variation
associated with changes in the atomic number is also small. In particular,
the atomic volumes of the transition metals in periods 3--5 (PO. 3--5)
are smaller than those of typical elements. Furthermore, the elements with the
minimum atomic volume correspond to those near the center of transition
metals in the periodic table, that is, those with \emph{d}-electron
numbers in the range of 6--8. The $V_{300} /V_{\rm 0amb}$ ratios of the
transition metals in periods
3--5 are also large and incompressible compared to typical elements and
exhibit a dome-shaped systematic atomic number dependence.
%Notably, even transition metals with a close-packed structure are incompressible.
%However, the group 4--6 elements Ti, Zr, Hf, V, Nb, Ta, Cr, Mo, and W,
%having a bcc structure among transition metals, are relatively easily
%compressed and have a small volume ratio of $V_{300}/V_{\rm 0amb}$, and
%their \emph{N} values are also small.
Notably, even transition metals of Cu, Ag, and Au with relatively small $K_0$
and a close-packed structure are incompressible in this pressure range.
On the other hand, the group 4–6 elements Ti, Zr, Hf, V, Nb, Ta, Cr, Mo,
and W, having a bcc structure among transition metals, are relatively
easily compressed and have a small volume ratio of $V_{300} /V_{\rm 0amb}$, and
their $N$ values are also smaller those with a close-packed structure. 
In addition, many elements exhibited \emph{N} values approximated four,
and elements with relatively small $V_{200}$ and $V_{300}$ tended to
have large \emph{N} values.
High-pressure metallic phases of Mg, P, and S of PO. 2 behave as a
simple metal with a bcc structure or a similar structure with pseudo-8
coordination, and their \emph{N} values are small approximately 2.5.

As mentioned earlier, the $P$--$V$ relationship of most
elements can be approximated by formula (3) using the coefficient
\emph{N} in the range of 200--300 GPa. As shown in Table I, the standard
deviation of \emph{N} values was within 0.5\%. Therefore, if the bulk
modulus at pressure \emph{P} is $K_{\rm p}$, the following
formula holds:

\begin{equation}
N = -d\log(P)/d\log(V) = -(V/P)\cdot(dP/dV) = K_{\rm P}/P
\end{equation}

Thus, $K_{\rm p}$ in this region can be estimated from the
following equation:

\begin{equation}
  K_{\rm P} = N \cdot P 
\end{equation}  

The parameter $N$ is also an indicator of the bulk modulus in this pressure range.
If $K_{\rm P}$ is first approximated as
$K_{\rm P} = K_{0} +K'_{0}\cdot P$ ($K_{0}, K'_{0}$ are constant),
the following relationship holds true:

\begin{equation}
K'_{0} = dK_{\rm P}/dP = N,
\end{equation}
which implies, $N$ corresponds to $K'_{0}$. For transition metals, the
$K'_{0}$ values obtained from the
experiments and $N$ values showed good correspondence, indicating
that this relationship can be applied. Thus, the $P$--$V$
relationship can be described simply in the multi-megabar region of
200--300 GPa. However, this discussion is limited to the pressure range
200--300 GPa and cannot be applied when a phase transition is involved
in this pressure range.

The \emph{P}--\emph{V} relations for the molecular solids
H\textsubscript{2}, He, Ne, N\textsubscript{2},\cite{ref58}
O\textsubscript{2}, Ar, Kr, and Xe are shown in Fig. 3. The volume
values for H, N, and O correspond to H\textsubscript{2},
N\textsubscript{2}, and O\textsubscript{2}, respectively. The
solidification pressure of each solid were indicated by arrows. Solid
H\textsubscript{2} and He have \emph{N} values of 2.27 and 2.46,
respectively, and are more compressible than other molecules whereas
solid Ne, N\textsubscript{2},\cite{ref58} O\textsubscript{2},
Ar, Kr, and Xe have similar \emph{N} values in the range of 3.3--3.5.
This may be because the wave functions of the 2\emph{s} and 3\emph{s}
electrons occupying the outer orbitals of O\textsubscript{2},
N\textsubscript{2}, and Ne, Ar, Kr, and Xe molecules are required to be
orthogonal to the wave function of the 1\emph{s} electron, but the
1\emph{s} electron cloud surrounding the H\textsubscript{2} and He
molecules do not have this orthogonality constraint.\cite{ref59}

Elements with close-packed structures (fcc and hcp) exhibit relatively
large \emph{N} values. This is particularly noticeable for transition
metals, where the \emph{N} value reaches 4.5--6. This is because the
\emph{d} electrons localized around the nucleus are a source of large
cohesive energy. Therefore, by further compressing a state in which the
electron density is already high, the volume reduction seems to cause a
rapid increase in the total electron kinetic energy.

Focusing on the individual elements, for Mn, a phase transition to the
hcp structure is proposed at approximately 165 GPa; therefore, measuring
the EOS of the hcp phase is necessary. Notably, the
\emph{V}\textsubscript{200} and \emph{V}\textsubscript{300} values of Hf
are smaller than those of other 3\emph{d} transition metals.

We were interested in determining why transition metals have large
\emph{N} values of 4.5--6. In a solid, under ultrahigh pressure
(200--300 GPa), the following quantum theoretical effects are expected
to play a role: (1) the Pauli exclusion principle, (2) orthogonality
constraints with core electrons, (3) relativistic effects for heavy
elements, and (4) the correlation of \emph{d}--electrons (although
magnetism disappears under ultrahigh pressure). These quantum
theoretical effects must be closely related to the total energy
\emph{E}\textsubscript{in}. Understanding the physical features of an
extremely high-density state of matter using quantum theory is important
for comprehending the behavior of materials under ultrahigh-pressure
conditions.

Finally, in the quantum theory of condensed materials, pressure is
defined as the volumetric derivative of the total energy
$E_{\rm in}$ (internal energy at temperature ($T) = 0$ K): $-P = dE_{\rm in}/dV$.
Namely, in this pressure range, $E_{\rm in}$ is expressed as
an $N-1$ order function of $V$: $E_{\rm in} \propto 1/V^{({\rm N}-1)}$.
$E_{\rm in}$ is the sum of the total kinetic energy of the electrons
$E_{k}$, electrostatic interaction $E_{\rm es}$, and exchange-correlation interaction,
$E_{\rm xc}$. We also analyzed the EOS of transition
metals obtained from previously reported density functional theory
calculations. The results obtained are in good agreement with the
results of this analysis. However, because we have not yet examined the
calculation methods and results in detail, we refrain from showing these
analytical results and comparing them with the experimental results.

\section{IV. Conclusion}

The current study demonstrated that for nearly all elementals, \emph{P}
could be very well approximated as an \emph{N}--order function of
\emph{V} in the multi-megabar pressure region between 200 and 300 GPa,
allowing us to easily evaluate the contribution of the volume change to
pressure, \emph{i.e}., the total energy of the system. We were able to
discuss the compressibility of elements systematically. However,
experimental data in this pressure range are still insufficient, and the
current study also suggests the need for further experimental and
theoretical research on elements in the pressure range. We hope that the
results obtained in this analysis will serve as basic data for future
ultrahigh-pressure research.

\section{SUPPLEMENTARY MATERIAL}

See the Supplementary Material for the comparison among equations of state
of Pt proposed by previous studies and the \emph{P}--\emph{V} curve for
the revised EOS of bcc--Hf using the Vinet equation.

\section{DATA AVAILABILITY}

The data that support the findings of this study are available from the
corresponding author upon reasonable request.

%\section{References}
%\bibliography{reference01.bib}

\section{Figure Captions}

FIG. 1. (color online) $\log(V)-\log(P)$ plots for (a)
Ni,\cite{ref9} (b) Pt,\cite{ref12,ref13,ref14} and (c)
Bi.\cite{ref15} Insets present results of the linear
approximation in a pressure range between 200 and 300 GPa.

FIG. 2. (color online) Atomic number change of (a)
\emph{V}\textsubscript{200}, \emph{V}\textsubscript{300}, (b)
\emph{V}\textsubscript{300}/\emph{V}\textsubscript{\rm 0amb} and (c)
\emph{N}-values for elements. Elements with \emph{P}\textsubscript{max}
higher than $\sim$200 GPa are shown in blue symbols. The
recalculated results for Hf, Ta, and Pt are shown with red symbols.

FIG. 3. (color online) \emph{P}--\emph{V} relation for molecular solids:
H\textsubscript{2}, He, Ne, N\textsubscript{2},\cite{ref58}
O\textsubscript{2}, Ar, Kr, and Xe. The values of volume for H, N, and O
correspond to a single molecule: H\textsubscript{2}, N\textsubscript{2},
and O\textsubscript{2}. Arrows show the solidification pressure for each
solid.

TABLE I. EOS data for elements and their analysis results: atomic No.,
element symbol, maximum pressure (P\textsubscript{max}, GPa), atomic
volume at ambient pressure (\emph{V}\textsubscript{0}, \AA\textsuperscript{3}),
bulk modulus (\emph{K}\textsubscript{0}, GPa),
pressure derivative of \emph{K}\textsubscript{0}
(\emph{K}'\textsubscript{0}), EOS formula,
Vinet (V), Birch-Murnaghan (BM), and AP1 (polynomial, see Ref. 37),
$N$ value ($P \propto 1/V^N$),
atomic volumes at 200 GPa and 300 GPa (\emph{V}\textsubscript{200}
and \emph{V}\textsubscript{300},\AA\textsuperscript{3}),
atomic volume of ambient pressure phase under ambient condition
(\emph{V}\textsubscript{0amb}, \AA\textsuperscript{3}),
volume ratio (\emph{V}\textsubscript{300}/\emph{V}\textsubscript{0amb}),
structure/phase, and reference No.
The column of structure/phase shows the high-pressure structure or phase
of the EOS based on the references (see the references for the structure
notation and details). 
The values of volume for hydrogen, nitrogen, and oxygen correspond
to a H\textsubscript{2}, N\textsubscript{2}, and O\textsubscript{2}
molecule. The value of \emph{V}\textsubscript{0amb} for rare gases,
hydrogen, nitrogen, and oxygen are correspond to
\emph{V}\textsubscript{0} in this table.

\begin{table*}[t]
  \centering
  \begin{longtable}{cccccccccccccc}
    \hline \hline
   Atomic No.
  & Atom
  & \emph{P}\textsubscript{max}
  & \emph{V}\textsubscript{0}
  & \emph{K}\textsubscript{0}
  & \emph{K}\textquotesingle{}\textsubscript{0}
  & EOS
  & \emph{N}, \emph{V}\textsuperscript{-\emph{N}}
  & \emph{V}\textsubscript{200}
  & \emph{V}\textsubscript{300}
  & \emph{V}\textsubscript{0amb}
  & \emph{V}\textsubscript{300}/ \emph{V}\textsubscript{0amb}
  & Structure /phase
  & Ref. \\
%\midrule\noalign{}
%\bottomrule\noalign{}
   %\endlastfoot
   \hline
1 & H & 190 & 42.246 & 0.110(6) & 7.36(7) & V & 2.273(3) & 3.6 & 3.01 &
42.25 & 0.071 & hcp & 2 \\
2 & He & 58 & 22.79 & 0.225 & 7.35 & V & 2.461(3) & 3.38 & 2.20 & 22.79
& 0.097 & fcc-hcp & 19 \\
4 & Be & 171 & 8.066 & 97.2(2.5) & 3.61(7) & BM & 3.05(1) & 4.19 & 3.66
& 8.07 & 0.454 & hcp & 17 \\
4 & Be & - & 8.066 & 96.59(5) & 3.758(3) & V & 2.99(1) & 4.18 & 3.64 &
8.07 & 0.451 & - & revised \\
6 & C & 80 & 5.669 & 444.5 & 4.18 & V & 5.29(3) & 4.36 & 4.03 & 5.669 &
0.711 & diamond & 20 \\
8 & O & 96 & 36.73 & 1.79(2) & 8.06(3) & V & 3.474(4) & 9.80 & 8.72 &
36.73 & 0.237 & $\varepsilon$-O\textsubscript{2} & 21 \\
10 & Ne & 208 & 22.234 & 1.07(2) & 8.40(3) & V & 3.465(4) & 5.41 & 4.81
& 22.23 & 0.216 & fcc & 22 \\
12 & Mg & 158 & 19.95 & 68.7 & 3.47 & V & 2.55(1) & 8.89 & 7.59 & 23.22
& 0.327 & bcc & 23 \\
13 & Al & 333 & 16.597 & 72.7 & 4.83 & V & 3.44(1) & 8.67 & 7.71 & 16.6
& 0.464 & fcc-hcp & 24 \\
14 & Si & 248 & 14.329 & 82 & 4.22 & BM & 3.654(8) & 7.62 & 6.82 & 20.02
& 0.341 & hcp & 25 \\
15 & P & 340 & 10.661 & 254(16) & 4.34(44) & V & 2.71(1) & 8.01 & 7.14 &
37.8 & 0.378 & sh & 26 \\
16 & S & 255 & 9.25 & 495(12) & 3.91(46) & V & 2.66(2) & 8.33 & 7.44 &
25.73 & 0.289 & $\beta$-Po & 27 \\
18 & Ar & 248 & 37.45 & 2.65 & 7.56(2) & v & 3.358(4) & 10.37 & 9.20 &
37.45 & 0.246 & fcc & 28 \\
21 & Sc & 300 & 15 & 98.5(1.0) & 5.28(4) & V & 3.91(1) & 8.71 & 7.95 &
24.79 & 0.32 & hexgonal & 29 \\
22 & Ti & 290 & 15.97(4) & 125(2) & 3.46(6) & BM & 3.20(2) & 8.9 & 7.84
& 17.74 & 0.442 & delta-bcc & 6 \\
23 & V & 290 & 13.87(2) & 156(6) & 4.3(1.6) & V & 4.12(2) & 8.67 & 7.86
& 13.87 & 0.567 & bcc-rhombo. & 7 \\
24 & Cr & 133 & 12.04(1) & 182(1) & 5.10(4) & V & 4.48(2) & 8 & 7.31 &
12.04 & 0.607 & bcc & 30 \\
25 & Mn & 190 & 12.66 & 158 & 4.6 & BM & 4.41(1) & 8.1 & 7.39 & 14.14 &
0.522 & $\alpha$-Mn & 31 \\
26 & Fe & 355 & 11.22(3) & 159.3(1.0) & 5.86(4) & V & 5.14(2) & 7.47 &
6.87 & 11.75 & 0.585 & hcp & 9 \\
27 & Co & 210 & 10.34 & 224 & 5.8 & V & 5.67(2) & 7.37 & 6.85 & 11.01 &
0.623 & fcc & 32 \\
28 & Ni & 370 & 10.9376(4) & 173.5(1.4) & 5.55(5) & V & 4.72(2) & 7.23 &
6.71 & 10.94 & 0.613 & fcc & 9 \\
29 & Cu & 144 & 11.81 & 132.4(1.4) & 5.32(6) & V & 4.26(2) & 7.4 & 6.77
& 11.81 & 0.573 & fcc & 33 \\
30 & Zn & 257 & 15.12(6) & 63(2) & 5.4(3) & BM & 4.237(6) & 8.22 & 7.47
& 15.12 & 0.494 & hcp & 8 \\
31 & Ga & 100 & 18.23(19) & 55(5) & 4.7(5) & BM & 3.806(5) & 9.13 & 8.21
& 19.6 & 0.419 & fcc & 34 \\
32 & Ge & 258 & 14.10(2) & 313(2) & 3.27(4) & BM & 3.85(2) & 9.81 & 8.83
& 22.64 & 0.39 & hcp & 8 \\
33 & As & 247 & 16.25(3) & 182(1) & 3.67(3) & BM & 3.668(8) & 10.15 &
9.09 & 21.51 & 0.423 & bcc & 8 \\
34 & Se & 320 & 15.52(9) & 290(6) & 3.20(7) & BM & 3.61(2) & 10.55 &
9.43 & 27.27 & 0.346 & bcc & 8 \\
35 & Br & 80 & 32.73 & 14.1 & 4.6 & BM & 3.350(3) & 11.06 & 9.83 & 42.54
& 0.231 & bct & 35 \\
36 & Kr & 140 & 37.47 & 3.03 & 7.24 & V & 3.243(4) & 10.28 & 9.07 &
37.47 & 0.242 & fcc-hcp & 36 \\
37 & Rb & 264 & 92.9(4) & 0.15(3) & 7.66(15) & AP1 & 3.166(2) & 12.28 &
10.81 & 92.90 & 0.116 & tI4-oC16 & 37 \\
39 & Y & 177 & 22.45 & 108 & 3.4 & BM & 2.72(1) & 11.65 & 10.03 & 32.9 &
0.305 & $C2/m$ & 38 \\
40 & Zr & 142 & 22.6 & 93(1) & 3.20(9) & V & 2.59(1) & 10.89 & 9.31 &
23.29 & 0.4 & bcc & 39 \\
41 & Nb & 134 & 18.008 & 151(3) & 4(1) & BM & 3.18(1) & 10.73 & 9.46 &
18.01 & 0.525 & bcc & 40 \\
42 & Mo & 410 & 15.58(1) & 262(4) & 4.55(16) & V & 4.60(2) & 10.95 &
10.03 & 15.58 & 0.644 & bcc & 3 \\
43 & Tc & 67 & 14.31 & 309(7) & 4.7(4) & V & 4.96(2) & 10.425 & 9.606 &
14.31 & 0.671 & hcp & 41 \\
44 & Ru & 151 & 13.565(2) & 319.1(9) & 4.40(2) & V & 4.78(1) & 9.858 &
9.057 & 13.565 & 0.668 & hcp & 42 \\
45 & Rh & 191 & 13.764(2) & 257(2) & 5.44(8) & V & 5.18(2) & 9.9 & 9.15
& 13.76 & 0.654 & fcc & 43 \\
46 & Pd & 182 & 14.669(18) & 189.3(30) & 5.473(63) & V & 4.78(2) & 9.98
& 9.15 & 14.67 & 0.624 & fcc & 44 \\
47 & Ag & 165 & 17.051(8) & 99.0(4) & 6.19(3) & V & 4.56(2) & 10.46 &
9.57 & 17.05 & 0.561 & fcc & 15 \\
48 & Cd & 174 & 21.578 & 42(1) & 6.5(2) & BM & 4.42(1) & 11.31 & 10.32 &
21.58 & 0.478 & hcp & 45 \\
49 & In & 250 & 26.18(1) & 32.7(10) & 6.28(25) & BM & 4.234(6) & 12.83 &
11.66 & 25.97 & 0.449 & dis. fcc & 5 \\
50 & Sn & 194 & 13.96(6) & 652(20) & 4 & BM & 4.026(4) & 13.15 & 11.9 &
27.04 & 0.44 & hcp & 46 \\
51 & Sb & 253 & 27.95(8) & 36.5(5) & 6.28(5) & V & 3.853(8) & 13.68 &
12.31 & 34.87 & 0.353 & bcc & 5 \\
52 & Te & 330 & 23.95(14) & 121(3) & 4.54(8) & V & 3.660(9) & 14.05 &
12.57 & 34.01 & 0.37 & fcc & 4 \\
53 & I & 276 & 31.25(7) & 30.38(13) & 6.13(1) & V & 3.660(9) & 14.41 &
12.91 & 40.9 & 0.316 & fcc & 47 \\
54 & Xe & 200 & 63.073 & 4.3(3) & 4.9(1) & BM & 3.309(1) & 15.77 & 13.95
&  & 0.221 & hcp & 48 \\
57 & La & 140 & 37.5 & 14.5(1) & 5 & BM & 3.549(3) & 13.53 & 12.08 &
37.19 & 0.325 & dis. fcc & 49 \\
72 & Hf & 120 & 21.5(2) & 109.8(5.2) & 3.10(5) & BM & 2.31(1) & 10.52 &
7.81 & 22.3 & 0.35 & bcc & 50 \\
72 & Hf & - & 15.32 & 279.28(8) & 2.079(5) & V & 2.48(1) & 10.82 & 9.18
& 22.3 & 0.412 & - & revised \\
73 & Ta & 90 & 18.035 & 197.0(3.5) & 3.39(15) & V & 3.38(2) & 11.23 &
9.96 & 18.04 & 0.552 & bcc & 13 \\
73 & Ta & - & 18.035 & - & - & V & 3.63(2) & 11.4 & 10.17 & 18.04 &
0.564 & - & revised \\
74 & W & 155 & 15.862 & 295.2(3.9) & 4.32(11) & V & 4.60(2) & 11.32 &
10.79 & 15.86 & 0.598 & bcc & 13 \\
75 & Re & 144 & 14.733 & 352.6 & 4.56 & V & 5.08(2) & 10.96 & 10.12 &
14.73 & 0.687 & hcp & 51 \\
76 & Os & 207 & 14.01 & 444 & 3.99 & BM & 5.31(2) & 10.75 & 9.96 & 14.01
& 0.711 & hcp & 52 \\
77 & Ir & 159 & 14.153(3) & 351(3) & 5.29(9) & V & 5.72(1) & 10.75 &
10.04 & 14.15 & 0.71 & fcc & 53 \\
78 & Pt & 800 & 15.102 & 259.7(2) & 5.839(3) & V & 5.48(3) & 10.99 &
10.21 & 15.1 & 0.676 & fcc & 14 \\
78 & Pt & 660 & 15.1 & 278 & 5.61 & V & 5.43(2) & 11.057 & 10.262 &
15.102 & 0.68 & fcc & 12 \\
78 & Pt & 100 & 15.095 & 273.6(2.0) & 5.23(8) & V & 5.14(2) & 10.91 &
10.09 & 15.1 & 0.668 & fcc & 13 \\
79 & Au & 222 & 16.929 & 170.9(2) & 5.880(5) & V & 4.94(2) & 11.44 &
10.53 & 16.93 & 0.622 & fcc & 14,54 \\
80 & Hg & 197 & 16.84(4) & 251(9) & 4.5(2) & V & 4.02(1) & 12.18 & 11.01
& 24.5 & 0.449 & hcp & 55 \\
81 & Tl & 125 & 28.2(2) & 48(4) & 4 & BM & 3.175(4) & 12.52 & 11.02 &
28.8 & 0.383 & fcc & 56 \\
82 & Pb & 238 & 29.8 & 39.9 & 6.13 & V & 3.82(1) & 14.75 & 13.27 & 30.33
& 0.438 & bcc & 57 \\
83 & Bi & 332 & 31.44(6) & 38.20(29) & 6.229(32) & V & 3.856(6) & 15.51
& 13.96 & 34.44 & 0.405 & bcc & 15 \\
\hline
\hline
\end{longtable}
\end{table*}

\begin{figure}
  \includegraphics[width=12cm]{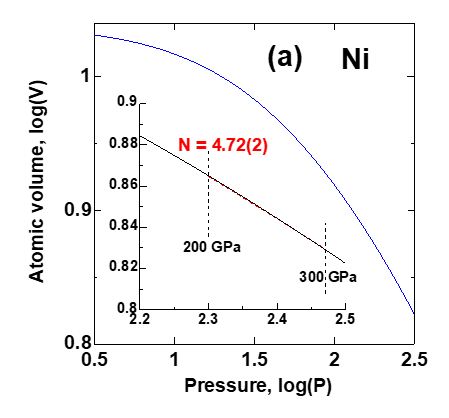}
\end{figure}

\begin{figure}
\includegraphics[width=12cm]{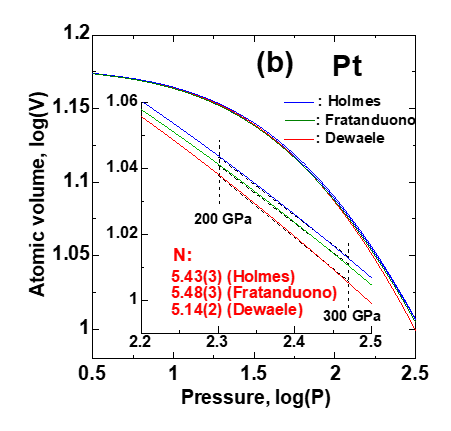}
\end{figure}

\begin{figure}
\includegraphics[width=12cm]{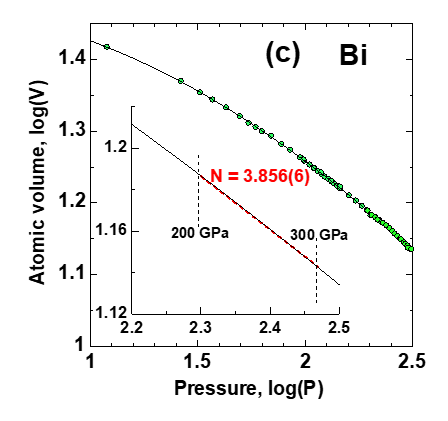}
\end{figure}

\begin{figure}
  \includegraphics[width=12cm]{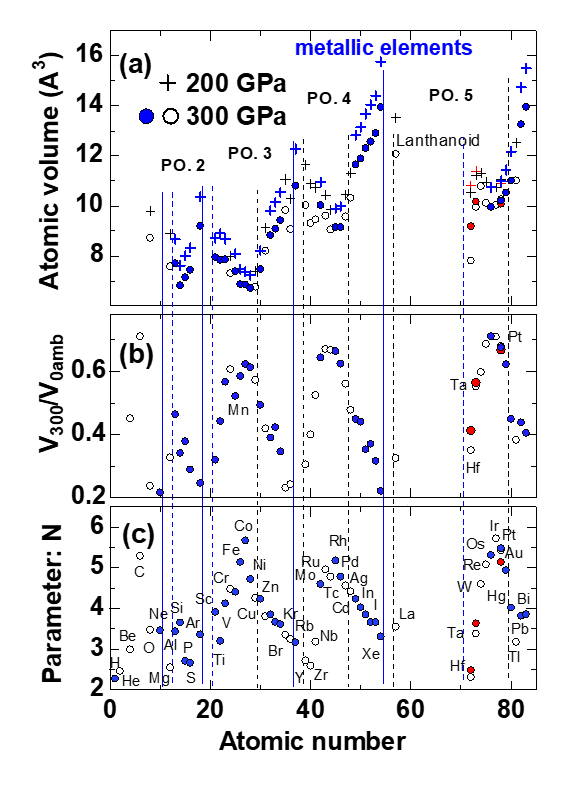}
\end{figure}

\begin{figure}
\includegraphics[width=12cm]{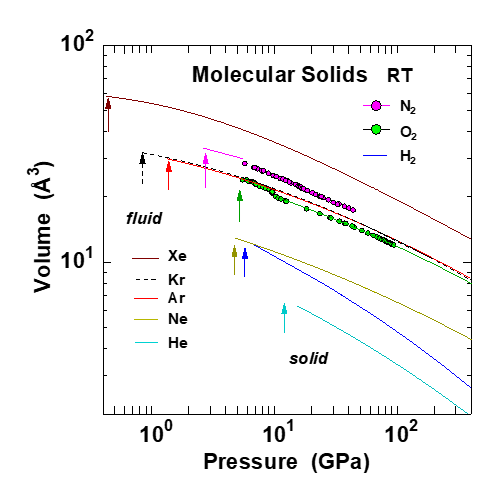}
\end{figure}

%\begin{quote}
%\includegraphics[width=4.66667in,height=4.3886in]{media/image4.emf}
%\end{quote}

Fig. 1. The $\log(V)-\log(P)$ plots for (a) Ni\cite{ref9},
(b) Pt\cite{ref12,ref13,ref14} and (c) Bi\cite{ref15}. Insets present
results of the linear
approximation in a pressure range between 200 and 300 GPa.

Fig. 2. Atomic number change of (a) \emph{V}\textsubscript{200},
\emph{V}\textsubscript{300}, (b)
\emph{V}\textsubscript{300}/\emph{V}\textsubscript{0amb} and (c)
\emph{N}-values for metallic elements. Elements with
\emph{P}\textsubscript{max} higher than about 200 GPa are shown in blue.
The recalculated results for Hf, Ta and Pt are shown with red symbols.

%\textbf{　}\includegraphics[width=5.18367in,height=5.19587in]{media/image6.emf}

Fig. 3. $P$--$V$ relation for molecular solids: H\textsubscript{2}, He, Ne,
N\textsubscript{2}\cite{ref57}, O\textsubscript{2}, Ar, Kr, and Xe. The
values of volume for H, N, and O correspond to one molecule:
H\textsubscript{2} and N\textsubscript{2}, and O\textsubscript{2}.
Arrows show the solidification pressure for each solid.

\end{document}